\PassOptionsToPackage{table}{xcolor}
\documentclass[PCJ,Unicode,screen]{cedram}

\usepackage[natbib=true,mincitenames=1,maxcitenames=2,maxbibnames=20,minbibnames=20,uniquelist=false,uniquename=false,style=pcj]{biblatex}
\usepackage{lipsum}
\usepackage{setspace}
\usepackage{amsmath}
\usepackage{enumitem}
\usepackage{tabularx}
\usepackage{amssymb}
\usepackage[utf8]{inputenc}
\usepackage{makecell}
\usepackage{graphicx}
\usepackage{lineno}
\usepackage{wasysym}
\usepackage{tikz}
\usepackage{xurl}
\usepackage{afterpage}
\nolinenumbers

\newcommand{\clr}{\textrm{CLR}}

\newcommand{\review}[1]{\textcolor{black}{#1}}

\providecommand*{\noopsort}[1]{}

\DOI{10.48550/arXiv.2502.07366} 
\PCIcorresp{solene.pety@inrae.fr}

\title[Hologenomic data simulation]{ Simulating transgenerational hologenomes under selection with RITHMS}
 \author{\firstname{Solène} \lastname{Pety}}
 \address{Université Paris-Saclay, INRAE, GABI, 78350, Jouy-en-Josas, France}
 \address{Université Paris-Saclay, INRAE, MaIAGE, 78350, Jouy-en-Josas, France}
 \email[S. Pety]{solene.pety@inrae.fr}
 \author{\firstname{Ingrid} \lastname{David}}
 \address{Université de Toulouse, INRAE, ENVT, GenPhySE, 31326, Castanet-Tolosan, France}
 \email[I. David]{ingrid.david@inrae.fr}
 \author{\firstname{Andrea} \lastname{Rau}}
 \address{Université Paris-Saclay, INRAE, GABI, 78350, Jouy-en-Josas, France}
 \email[A. Rau]{andrea.rau@inrae.fr}
 \author{\firstname{Mahendra} \lastname{Mariadassou}}
 \address{Université Paris-Saclay, INRAE, MaIAGE, 78350, Jouy-en-Josas, France}
 \email[M. Mariadassou]{mahendra.mariadassou@inrae.fr}

 \begin{abstract}
     A holobiont is made up of a host organism together with its microbiota. In the context of animal breeding, the holobiont can be viewed as the single unit upon which selection operates. Therefore, integrating microbiota data into genomic prediction models may be a promising approach to improve predictions of phenotypic and genetic values. Nevertheless, there is a paucity of hologenomic transgenerational data to address this hypothesis, and thus to fill this gap, we propose a new simulation framework. Our approach, an R Implementation of a Transgenerational Hologenomic Model-based Simulator (RITHMS) is an open-source package. It builds upon simulated transgenerational genotypes from the Modular Breeding Program Simulator (MoBPS) package and incorporates distinctive characteristics of the microbiota, notably vertical and horizontal transmission as well as modulation due to the environment and host genetics. In addition, RITHMS can account for a variety of selection strategies and is adaptable to different genetic architectures. We simulated transgenerational hologenomic data using RITHMS under a wide variety of scenarios, varying heritability, microbiability, and microbiota transmissibility. We found that simulated data accurately preserved key characteristics across generations, notably microbial diversity metrics, exhibited the expected behavior in terms of correlation between taxa and of modulation of vertical and horizontal transmission, response to environmental effects and the evolution of phenotypic values depending on selection strategy. Our results support the relevance of our simulation framework and illustrate its possible use for building a selection index balancing genetic gain and microbial diversity and for evaluating the impact of partially observed microbiota data. RITHMS is an advanced, flexible tool for generating transgenerational hologenomes under selection that incorporate the complex interplay between genetics, microbiota and environment.
 \end{abstract}
\begin{document}

\maketitle

\onehalfspacing
\section*{Introduction}


An individual host organism and its associated microbiota together constitute a holobiont, a complex biological system characterized by symbiotic interactions \citep{ghataoraRewiringHolobiontSystems2025,guerreroSymbiogenesisHolobiontUnit2013}. The concept of the holobiont was introduced several decades ago \citep{margulisSymbiosisSourceEvolutionary1991} and, although sometimes debated \citep{theisGettingHologenomeConcept2016}, it has gained increasing recognition as a relevant biological entity on its own, integrating both the host and its microbial partners. The holobiont concept has notably garnered attention in recent years in studies focused on the sustainable production of livestock species within a One Health framework \citep{evansHistoryOneHealth2014}. In particular, the complex dependencies between host genetics and microbiota, together with other environmental factors, have been shown to impact agriculturally important phenotypes such as digestive and feed efficiency \citep{liHostGeneticsInfluence2019, deruGeneticRelationshipsEfficiency2022, niuDynamicDistributionGut2015}, milk composition \citep{martinezboggioHostGeneticControl2022, brulinShortCommunicationBifidobacterium2024} and mastitis triggering \citep{derakhshaniInvitedReviewMicrobiota2018a}.

The co-evolution of the host genome and its microbiome under selective pressure suggest the potential interest of using the holobiont as a selection unit for animal breeding \\ \citep{zilber-rosenbergRoleMicroorganismsEvolution2008}. Since the end of the Second World War, selection has led to substantial gains in performance traits for livestock \citep{goldbergComparativeAnalysisSelection1991}, further strengthened in the past two decades by the widespread use of genomic data to estimate breeding values. These genomic prediction methods are most often based on linear models assuming additive genetic effects, allowing a decomposition of phenotypic variance into genetic and environmental components \citep{legarraBasesGenomicPrediction2018, lealGeneticsAnalysisQuantitative2001}. In this quantitative genetics framework, a phenotype is simply the measurable trait of interest (e.g. milk yield, growth, or disease resistance), often considered on a continuous scale.
Thanks to the relatively low cost of sequencing technologies, it is now straightforward to assay the microbiota using metabarcoding data in addition to host genomic data. However, jointly considering host genotypes and microbiota for selection is complicated in part by the multiple potential modes of transmission of the microbiota and in part by their interplay on the phenotype.
For example, in the first few moments of existence for mammals, maternal contact during delivery and nursing play a crucial role in establishing the initial microbiota through vertical transmission (\cite{rutayisireModeDeliveryAffects2016,cortes-maciasMaternalDietShapes2021,bruijningNaturalSelectionImprecise2022}). For non-mammalian vertebrates, such as chickens (\cite{shterzerVerticalTransmissionGut2023}), the maternal contribution is likely to be considerably weaker than in mammals. In stark contrast to genotypes, a fraction of the microbiota is acquired from the environment through horizontal transmission, and the microbiota continues to evolve throughout a host's life. In addition, both host genes and environmental factors influence the colonization, development, and function of the microbiota, which in turn contributes to host phenotypes.

Several recent studies have sought to evaluate the added value of hologenomic as compared to genomic selection (\cite{rossMetagenomicPredictionsMicrobiome2013a,weishaarSelectingHologenomeBreed2020,martinezboggioHostRumenMicrobiome2024a}). In large-scale studies that collect microbiome data on a large number of individuals, as is done in genomic studies, the microbiota is typically measured at a single time point (or at most a few). Consequently, what is analysed is not the entire temporal dynamic of microbiota, but the resulting community at a specific and relatively stable stage. Increasing the scale and scope of such studies raises a number of statistical and computational challenges with respect to the simultaneous integration of host genotypes and microbiota. Moreover, benchmarking studies to evaluate predictive hologenomic models require sufficiently large and fully paired transgenerational genomic and microbiota data, notably for the comparison of predicted breeding values to observed offspring phenotypes. Such experimental data are costly to acquire and can be impacted by biases (\emph{e.g}., fluctuating environmental conditions). 

Simulation therefore offers an efficient and cost-effective approach to assess the relevance of hologenomic prediction strategies based on host genotypes and a stable snapshot of the microbiota. 
Several tools for simulating transgenerational genotypes are well known and implemented in user-friendly software such as the Modular Breeding Program Simulator (MoBPS) 
(\cite{pookMoBPSModularBreeding2020}) or AlphaSim (\cite{fauxAlphaSimSoftwareBreeding2016}). These tools provide flexible and efficient implementations that allow for a wide range of breeding schemes under a variety of scenarios (\emph{e.g.} heritability) but do not integrate microbiota data. With respect to hologenomic data, other simulation approaches have focused on modeling the structure and dynamics of the microbiota, both for exploring breeding strategies, as well as integrating different types of data to account for complex microbiota-host genome interactions (\cite{perez-encisoOpportunitiesLimitsCombining2021,wirbelRealisticBenchmarkDifferential2024}). These methods all focus on the potential added value of the microbiota for selection, and thus generally focus on generating a stable snapshot rather than attempting to model and reproduce its dynamic throughout the lifetime of the host, even though this may result in a lack of power. This focus on a single, relatively stable time point is in line with current large-scale microbiome studies, which typically measure the microbiota at only one (or a few) sampling time.
Concepts such as transmissibility have also emerged, taking into account the transmissibility of non-genetic information and thus broadening the vision of inclusive heritability (\cite{davidUnifiedModelInclusive2019}). However, none of these hologenomic simulation approaches incorporate a transgenerational aspect. 

One recent exception for simulating a co-evolving genome and microbiota under selection is HoloSimR (\cite{casto-rebolloHoloSimRComprehensiveFramework2025}), which generates a fully synthetic set of genomic and microbiota data over multiple generations based on a user-provided population demographic history and pre-defined model of species abundance distribution using AlphaSimR \citep{gaynorAlphaSimRPackageBreeding2021} and mobsim \citep{mayMobsimPackageSimulation2018}, respectively.
    Although it constitutes a solid approach for simulating transgenerational hologenomic data, HoloSimR does have several limitations. First, real hologenomic datasets cannot be used to initialize simulations, and the base population is directly artificially generated under assumptions that may oversimplify the complex patterns within and between microbiota and genomic data. Second, it uses a single heritability value for all taxa and thus cannot reproduce the distribution of heritability values inherent to real holobionts, with some taxa being selected for and highly heritable whereas other are selected against and less heritable. Third, the impact of short- and long-term environmental perturbations on the microbiota, such as antibiotic treatments or diet changes, cannot be incorporated in the simulations.
    
    To address these limitations, we introduce an R Implementation of a Transgenerational Hologenomic Model-based Simulator (RITHMS). RITHMS is a flexible framework for simulating transgenerational hologenomic data that accounts for the specificities of microbiota transmission and covers the same range of breeding schemes as MoBPS, but under additional and more complex scenarios (heritability, microbiability, microbiota transmissibility, etc). Real genomic and microbiota data are used to construct a base population from which subsequent generations are derived. This work describes the general framework and strategy used for simulations, and shows that simulated data preserve key characteristics of real data. Finally, it demonstrates the usefulness of transgenerational hologenomic data simulated with RITHMS through a case study on mixed-objective selection, incorporating both phenotype values and microbial diversity, as well as a case study focusing on partially unobserved microbiota data.

\section*{Material and methods }

RITHMS directly incorporates the complexity of transgenerational hologenomic data in several ways (Figure \ref{figure:graphical_wf}): (1) it uses user-provided paired genomic and microbiota data to create a realistic base population from which successive non-overlapping generations are generated, (2) it takes into account the particularities of microbiota transmission (vertical and horizontal) and genetic modulation,  (3) it leverages the functionalities provided in  MoBPS (\cite{pookMoBPSModularBreeding2020}) to define complex genetic architectures and breeding selection steps using indices based on breeding values, microbial descriptors, or a combination of the two, and (4) it facilitates simulations under a variety of scenarios. 

RITHMS works with an initialization step of the base population followed by repeated steps (once per generation) of simulation for subsequent generations, summarized in Figure~\ref{figure:graphical_wf} and discussed in greater detail in the following. Key simulation parameters and notations are detailed in Table~\ref{table:notations}.

\begin{table}[]
\caption{Table of key parameters and notations for RITHMS}
\rowcolors{3}{gray!15}{white}
\centering
\small
\begin{tabularx}{\textwidth}{| c | X | c | c | c |} 
 \hline
 \rowcolor{gray!50} \textbf{Symbol} & \textbf{Definition} & \textbf{Dimensions} & \review{\makecell[l]{\textbf{Default}\\ \textbf{initialization}}} & \textbf{RITHMS parameter} \\ [0.5ex] 
 \hline
 $N$ & Number of individuals in the base population & $1 \times 1$ & (*) & \\
 $n_{\text{gen}}$ & Number of generations after the base population & $1 \times 1$ & 5 & \texttt{n\_gen} \\
 $n_{\text{ind}}$ & Number of individuals per generation & $1 \times 1$ & $N$ & \texttt{n\_ind}  \\
 \hline
 \rowcolor{gray!50} \multicolumn{5}{|l|}{Genotype parameters and notations}\\
 \hline
 $n_\text{g}$ & Number of SNPs & $1 \times 1$ & (*) &  \\
 $v$ & Number of genetic clusters & $1 \times 1$ & (*) &  \\
 $\boldsymbol{G}^{(0)}$ & Base population genotypes encoded as 0,1,2 & $n_\text{g} \times N$ & (*) &  \\
 $\boldsymbol{G}^{(t)}$ & Genotypes encoded as 0,1,2 & $n_\text{g} \times n_{\text{ind}}$ & & \\
 $\text{QTL}_\text{y}$ & Number of causative QTL on the phenotype & $1 \times 1$ & 100 & \texttt{qtn\_y} \\
 $\text{QTL}_\text{o}$ & Number of causative QTL on taxa abundances (per taxon) & $1 \times 1$ & \makecell{(*)  \\ $n_\text{g}*0.2 /v$} &  \\
 \hline
 \rowcolor{gray!50} \multicolumn{5}{|l|}{Microbiota parameters and notations}\\
 \hline
 $n_\text{b}$ & Number of taxa & $1 \times 1$ & (*) & \\
 $\lambda$ & Proportion of vertical transmission & $1 \times 1$ & $0.5$ & \texttt{lambda} \\
 $\boldsymbol{\beta}$ & QTL effects on taxa abundances & $n_\text{b} \times n_\text{g}$ & \review{$\sim \mathcal{N}(0, \sigma^2_{\beta})$} &  \\
 $\sigma_{\beta}$ & Standard deviation for non-null cluster- and taxon-specific genetic effects on taxa abundances & $1 \times 1$ & $0.1$ & \texttt{effect\_size}  \\
 $\sigma_m$ & Standard deviation value for microbiota noise & $1 \times 1$ & $0.1$ & \texttt{noise.microbiome}  \\
 $\boldsymbol{M}^{(0)}$ & Base population taxa counts & $n_\text{b} \times N$ & (*) &  \\
 $\boldsymbol{M}^{(t)}$ & Taxa abundances & $n_\text{b} \times n_\text{ind}$ & &  \\
  $\boldsymbol{B}^{(t)}$ & CLR-transformed relative abundance values for taxa of all individuals at generation $t$, $\clr(\boldsymbol{M}^{(t)})$ & $n_\text{b} \times n_\text{ind}$ & &  \\
  \review{$k$} & \review{Number of fixed effects} & \review{$1 \times 1$} & & \\
  \review{$\boldsymbol{\theta}$} & \review{Environmental effects on taxa abundances} & \review{$n_\text{b} \times k$} & &  \\
  \review{$\boldsymbol{E}^{(t)}$} & \review{Environmental factors} & \review{$n_\text{ind} \times k$} & &  \\
 $\text{OTU}_\text{g}$ & Percentage of taxa under genetic control  & $1 \times 1$ & $5\%$ & \texttt{otu\_g}\\
 $\text{OTU}_\text{y}$ & Percentage of causative taxa on phenotype & $1 \times 1$ & \texttt{otu\_g} & \\
 \hline
 \rowcolor{gray!50} \multicolumn{5}{|l|}{Phenotype parameters and notations}\\
 \hline
 $\boldsymbol{y}^{(t)}$ & Phenotype, $\boldsymbol{\alpha}^T \boldsymbol{G}^{(t)} + \boldsymbol{\omega}^T \boldsymbol{B}^{(t)} + \boldsymbol{\epsilon}_y^{(t)}$ & $n_\text{ind} \times 1$ & &  \\
 $\boldsymbol{\omega}$ & Taxa effects on phenotype & $1 \times n_\text{b}$ & \review{$\sim \Gamma(1.4, 3.8)$} & \\
 $\boldsymbol{\alpha}$ & QTL effects on phenotype & $1 \times n_\text{g}$ & \review{$\sim \Gamma(0.4, 5)$} & \\
  $h_\text{d}^2$ & Direct heritability, $\text{var}(\boldsymbol{\alpha}^T \boldsymbol{G}^{(t)}) / \text{var}(\boldsymbol{y}^{(t)})$ & $1 \times 1$ & (**) & \texttt{h2} \\
  $\boldsymbol{\omega}^T \boldsymbol{B}^{(t)}$ & Microbiota effect & $n_\text{ind} \times 1$ & & \\
 $b^2$ & Microbiability, $\text{var}(\boldsymbol{\omega}^T \boldsymbol{B}^{(t)}) / \text{var}(\boldsymbol{y}^{(t)})$ & $1 \times 1$ & (**) & \texttt{b2} \\
 $h^2$ & \makecell[l]{Total heritability,\\ $[ \text{var}(\boldsymbol{\alpha}^T \boldsymbol{G}^{(t)}) + \text{var}(\boldsymbol{\omega}^T \boldsymbol{\beta G}^{(t)})] \big/ \text{var}(\boldsymbol{y}^{(t)})$} & $1 \times 1$ & & \\ 
  $\textbf{BV}_t^{(t)}$ & \makecell[l]{Total breeding value, all genetic effects, \\ $\textbf{BV}_t^{(t)} = \textbf{BV}_m^{(t)} + \textbf{BV}_d^{(t)}$} & $n_\text{ind} \times 1$ & & \\
 $\textbf{BV}_m^{(t)}$ & \makecell[l]{Microbiota-mediated breeding value, \\ $\textbf{BV}_m^{(t)} = \boldsymbol{\omega}^T \boldsymbol{\beta G}^{(t)}$} & $n_\text{ind} \times 1$ & & \\
 $\textbf{BV}_d^{(t)}$ & Direct breeding value, $\textbf{BV}_d^{(t)} = \boldsymbol{\alpha}^T \boldsymbol{G}^{(t)}$ & $n_\text{ind} \times 1$ & & \\
 \hline
\end{tabularx}
\label{table:notations}
\begin{flushleft}
\footnotesize (*) = Calibrated from input data, (**) = Required parameters,  $^{(t)}$ indicates quantities pertaining to generation $t$
\end{flushleft}
\end{table}

\afterpage{
\begin{landscape}
    \begin{figure}
    \centering
    \vspace*{40pt}
    \includegraphics[width=700pt]{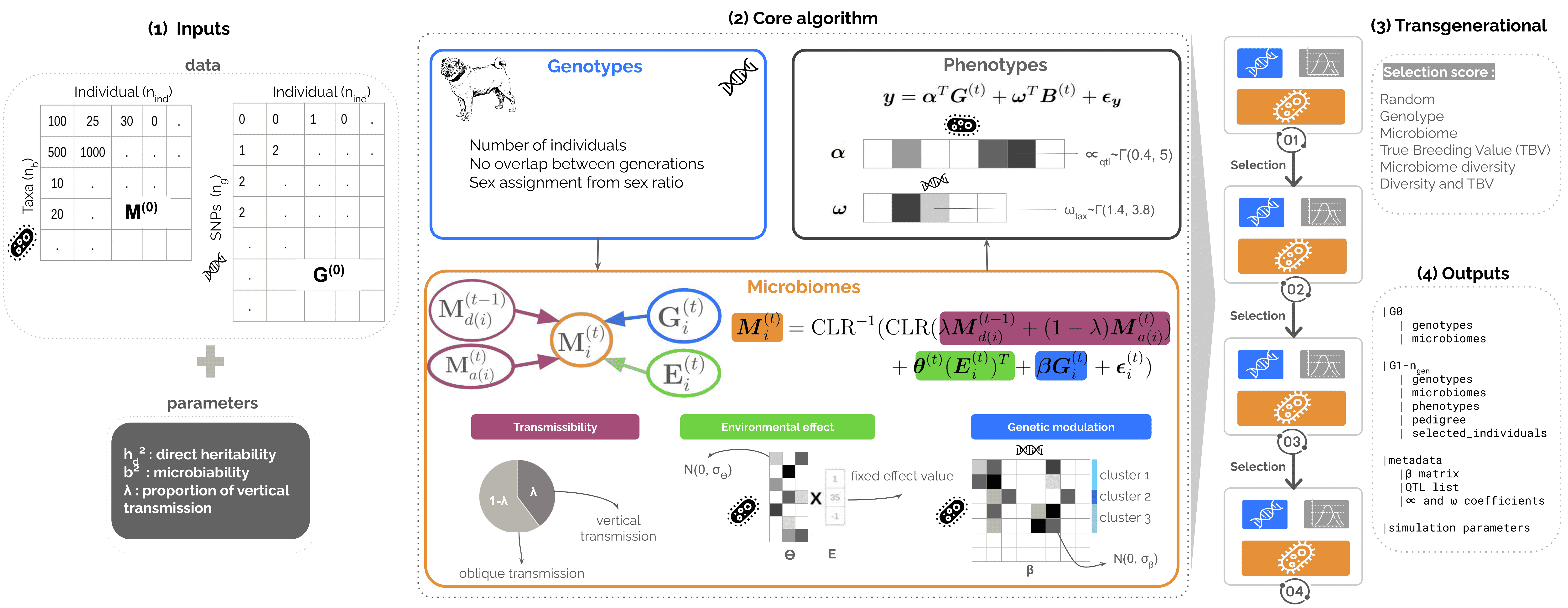}
    \caption{\label{figure:graphical_wf} Overview of RITHMS. (1) User-provided inputs include paired microbiota abundances and genotypes (encoded as 0/1/2) and the following required parameters: direct heritability $h^2_d$, microbiability $b^2$ and $\lambda$, which modulates the vertical versus horizontal transmission ratio. (2) For each simulated generation $t$, the genotypes and pedigree are generated using the MoBPS package \citep{pookMoBPSModularBreeding2020}. Microbiota are then constructed by first combining maternal and ambient microbiota in proportions $\lambda$ and $1 - \lambda$ respectively, and subsequently applying genetic and possibly environmental modulation. Genotypes and microbiota are then integrated to simulate the phenotypes of the generation using the recursive model of \citet{perez-encisoOpportunitiesLimitsCombining2021}. (3) To proceed with the next generation, 30\% of the males and 30\% of the females are selected, either randomly or based on a selection index chosen by the user. (4) The output of the package consists of a nested list including the simulation parameters and, for each generation $t$, the microbiota $\boldsymbol{M}^{(t)}$ and genotype $\boldsymbol{G}^{(t)}$ tables.}
    \end{figure}
\end{landscape}
}

\subsection*{Formatting of the base population}

The base population corresponds to user-provided paired genotype $\boldsymbol{G}^{(0)}$ and microbiota $\boldsymbol{M}^{(0)}$ data for $N$ individuals, with respectively $n_\text{g}$ SNPs and $n_\text{b}$ taxa. For the base population alone (G0), random matings among all individuals are used to generate the following generation (G1) using MoBPS. Exactly $n_{\text{ind}}$ individuals are generated according to the user-specified sex ratio (0.5 by default).

\subsubsection*{Genotype data}

Genotype data $\boldsymbol{G}^{(0)}$ should be coded as the number of alternative alleles at each variant for each individual (a $n_\text{g} \times N$ matrix). Genotypes are provided by the user and can therefore correspond to real data, ensuring realistic linkage disequilibrium (LD) and allelic frequency features in subsequent simulated generations. Sex chromosomes and sample meta-data beyond individual identifiers are not used for the simulations and are ignored if present. Each individual in the population is assigned as female or male according to a sex ratio parameter (set to 0.5 by default).

\subsubsection*{Microbiota data}
As for the base population genotypes, $\boldsymbol{M}^{(0)}$ can be based on real data, as long as they can be summarized as count table. Our motivating example and illustration are based on 16S rRNA metabarcoding data as they are the most common in breeding studies \citep{goodrichGeneticDeterminantsGut2016} but it could be equally applied to WGS data. A  $n_\text{b} \times N$ count matrix is expected, with the same $N$ individuals as those from $\boldsymbol{G}^{(0)}$. No modulation is applied to this initial microbiota as it is already considered to be under genetic influence.
The provided raw abundances of taxa are used to estimate compositions (\emph{i.e.} vectors of relative abundances) using an empirical Bayes approach to avoid zeroes, as a data-driven alternative to pseudocounts. To remove the compositionality constraint when incorporating genetic and environmental modulations, relative abundances are subsequently transformed using the centered log-ratio (CLR) from the \textit{compositions} R package \citep{vandenboogaartCompositionsUnifiedPackage2008},  
$\text{clr}(\mathbf{x}) = \left(\ln x_i - \frac{1}{n_b} \sum_{j=1}^{n_b} \ln x_j \right)_i$. This transformation \review{formally} includes a centering term such that the resulting values sum to zero \citep{gloorMicrobiomeDatasetsAre2017}. \review{However, backward transformation to relative abundances is performed using the inverse CLR function (\texttt{clrinv}), which is invariant to translation, such as centering. We can therefore omit it from our equations without loss of generality.} The empirical Bayes approach uses a Dirichlet prior  $\mathcal{D}(S\mathbf{p})$ with scale parameter $S$ and mean parameter $\mathbf{p}$. The latter is set to the population-level composition, estimated as the average of relative abundances across all individuals. In practice, the composition of an individual thus corresponds to a weighted average between its empirical composition and that of the mean population with proportions $\pi$ and $1 - \pi$ respectively. The default value $\pi = 0.75$ corresponds to a scale parameter $S$ set to a third of the sample total count.

\subsubsection*{Real hologenomic data used as a base population}
\label{sec:dataset}

To illustrate the functionality of RITHMS, a set of hologenomic data from a single line of $N=750$ pigs fed a conventional diet \citep{deruImpactHighfibreDiet2020} were used as a base population in this work. Individual pigs were genotyped using a 70K SNP GeneSeek GGP Porcine HD chip, and microbiota composition was analysed using the V3-V4 region of 16S rRNA gene (see \citet{deruGeneticRelationshipsEfficiency2022} for additional details on data acquisition and processing). 
We focused here on a subset of the first 5000 SNPs from the chip manifest as well as the 1845 taxa with a prevalence higher than 5\% after rarefaction (rarefied depth = 4100 reads).

\begin{figure}[t!]
\centerline{\includegraphics[width=400pt]{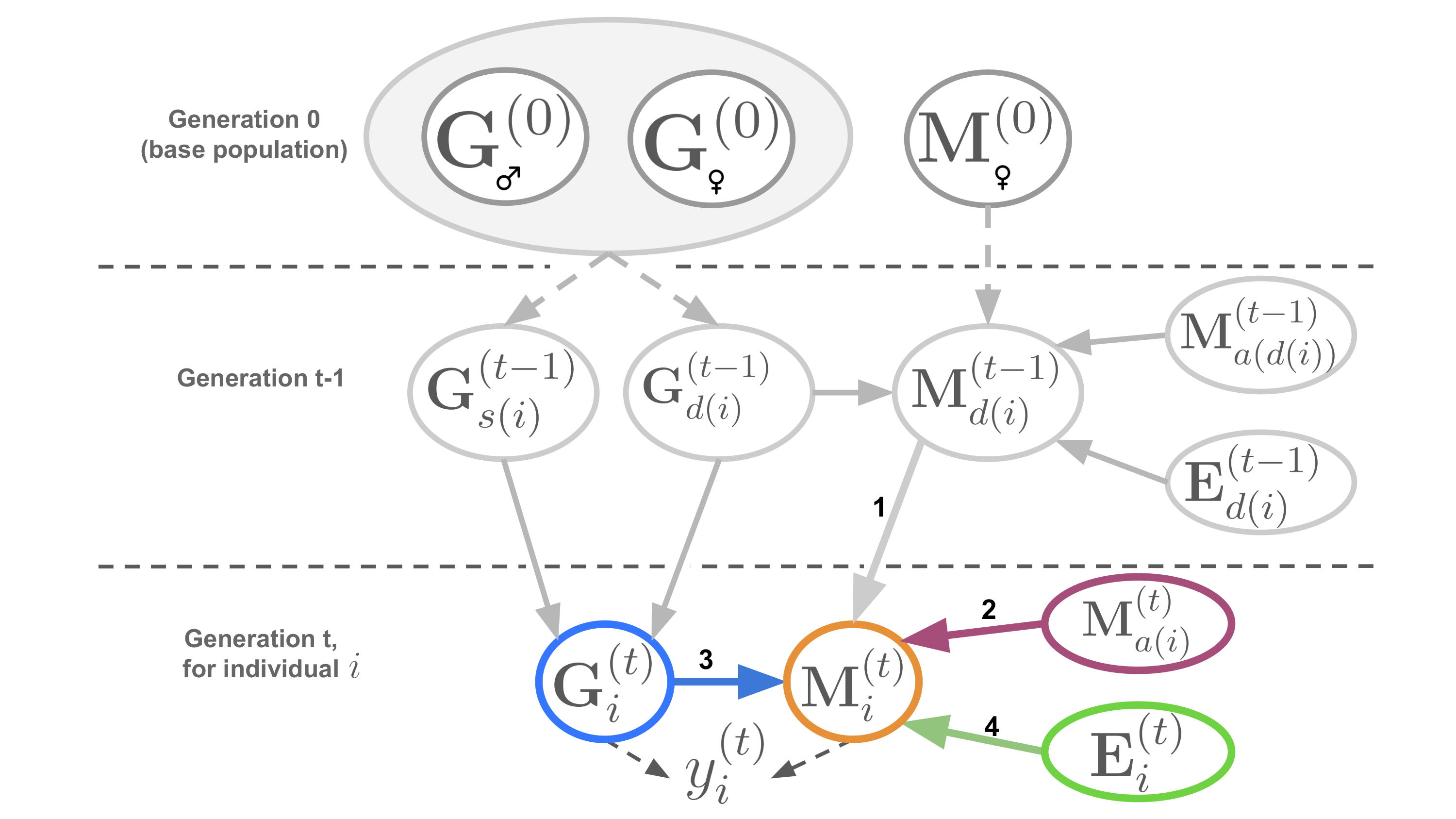}}
\caption{\label{figure:bubble_chart}Schematic illustration of transgenerational hologenomic simulations with RITHMS. The base generation is calibrated on user-provided data for sire genotypes $\boldsymbol{G}_{\male}^{(0)}$, dam genotypes $\boldsymbol{G}_{\female}^{(0)}$, and microbiota data from dams $\boldsymbol{M}_{\female}^{(0)}$. Genotypes are simulated using MoBPS \citep{pookMoBPSModularBreeding2020}. The sources contributing to the variability of taxa abundances of individual $i$ at generation $t$ are as follows: (1) vertical transmission from the individual's mother $\boldsymbol{M}_{d(i)}^{(t-1)}$, for example during delivery; (2) horizontal transmission of the individual-specific ambient microbiota $\boldsymbol{M}_{a(i)}^{(t)}$; (3) the host selective filter, through which the host's genotype $\boldsymbol{G}_i^{(t)}$ facilitates the colonization and establishment of certain microorganisms; and (4) individual-specific environmental effects $\mathbf{E}_i^{(t)}$, such as diet or treatment effects, modulating the microbiota composition. Phenotypes $\boldsymbol{y}_i^{(t)}$ are simulated according to a linear model, where the microbiota has a direct effect and the genome has both direct and microbiota-mediated effects \citep{perez-encisoOpportunitiesLimitsCombining2021}.}
\end{figure}

\subsection*{Simulation of subsequent generations}

\subsubsection*{Simulation of genotypes}

We simulate pedigrees and successive generations of genotype data using MoBPS based on the genotype data provided as input. By default, $n_{\textrm{gen}} = 5$ non-overlapping generations are simulated, each with $n_{\textrm{ind}} = 500$ individuals and a sex ratio of 0.5. Regardless of the simulation strategy employed, 30\% of females and 30\% of males at each generation are chosen to reproduce for the following generation, with selection either performed randomly or based on a user-specific selection score (see the \textit{Selection} section).

\subsubsection*{Simulation of microbiota}

As our goal is to evaluate the potential added value of hologenomic data in breeding schemes, we aim to simulate a stable snapshot of microbiota data that incorporates both genetic and environmental modulations.
Our simulation framework is based on the idea that the initial composition of an individual's microbiota can be partly inherited from its mother through vertical transmission as well as from the direct environment and is subsequently modulated by the host genotype and environmental factors prior to the snapshot (Figure \ref{figure:bubble_chart}).
\review{Therefore, we propose the following model for the microbiota abundances of individual $i$ at generation $t$ ($t = 1, \dots, n_{\text{gen}}$):}

\begin{align*}
\boldsymbol{M}^{(t)}_{i} =  \boldsymbol{\clr}^{-1}\left(\boldsymbol{\clr}\left(\lambda \boldsymbol{M}^{(t-1)}_{d(i)} + (1 - \lambda) \boldsymbol{M}^{(t)}_{a(i)} \right) \notag + \boldsymbol{\theta} (\boldsymbol{E}_i^{(t)})^T + \boldsymbol{\beta} \boldsymbol{G}_i^{(t)} + \boldsymbol{\epsilon}_i^{(t)}\right)
\end{align*}    
based on the following matrices, with their dimensions :
\begin{itemize}[itemsep=0pt,parsep=0pt]
\item $\boldsymbol{M}^{(t)}_{i}$ : taxa abundances in individual $i$ ($n_\text{b} \times 1$)
\item $\lambda$ : proportion of microbiota inherited via vertical transmission from the mother before modulation by selective filtering and random perturbations
\item $\boldsymbol{M}^{(t-1)}_{d(i)}$ : taxa abundances of the dam of individual $i$ ($n_\text{b} \times1$)
\item $\boldsymbol{M}^{(t)}_{a(i)}$ : Ambient taxa abundances for individual $i$ ($n_\text{b} \times1$)
\item $\boldsymbol{\theta}$ : environmental effects on taxa abundances ($n_\text{b} \times k$)
\item $\boldsymbol{E}_i^{(t)}$ : environmental factors for individual $i$ ($1 \times k$)
\item $\boldsymbol{\beta}$ : multiplicative effect of genotype on taxa abundances ($n_\text{b} \times n_\text{g}$)
\item $\boldsymbol{G}_i^{(t)}$ : genotype of individual $i$ ($n_\text{g} \times 1$)
\item $\boldsymbol{\epsilon}_i^{(t)} \sim \mathcal{N}(0,\,\sigma^2_m \mathbf{I}_{n_\text{b}})$: multivariate Gaussian white noise.
\end{itemize}

\paragraph{\textit{Ambient microbiota for each individual}}

We assume that no herd structure is considered here, that is, that individuals from the same generation live in similar conditions and are therefore exposed to the same sources of microorganisms.
As is common in breeding simulations, we further assume non-overlapping generations. To account for these constraints and avoid confounding horizontal and  direct non-maternal transmission, we propose a horizontal transmission that is conceptually similar to an oblique transmission between generations. Specifically, we propose a two-step process based on (1) an ambient microbiota that evolves slowly across generations, and (2) an individual-specific incorporation of this ambient microbiota.

First, to simulate a slowly-evolving ambient microbiota at generation $t$, we use the average composition of the previous generation, $\boldsymbol{\overline{M}}^{(t-1)}$. Second, we allow each individual to integrate these potential new communities in a different way, leading to the need to include inter-individual variability in the ambient microbiota composition while preserving the structure of the average composition of the previous generation. To introduce inter-individual variability, a random composition is sampled from a Dirichlet prior $\boldsymbol{M}_{r(i)}^{(t)} \sim \text{Dir}(\eta\boldsymbol{\overline{M}}^{(t-1)})$, where $\eta > 0$ (set to 25 by default) is the dispersion parameter calibrated to mimic the dispersion in the base population. Here, $\eta$ was set to 25 based on visual inspection and minimal differentiation of real and simulated compositions according to a PERMANOVA test (results not shown). Similar values of $\eta$ were found with other datasets \citep{perez-encisoOpportunitiesLimitsCombining2021, chaillouOriginEcologicalSelection2015}. Although this sampling results in compositions centered around $\boldsymbol{\overline{M}}^{(t-1)}$, extremely low abundances, corresponding to very large CLR-transformed values may occur,  thus overwhelming any possible modulation by noise or genetics. To regularize this sampling, in particular when $\lambda = 0$ (\emph{i.e.} no vertical transmission), we compute a weighted average between the sampled composition and the average composition of the previous generation $\boldsymbol{\overline{M}}^{(t-1)}$: 
\begin{equation*}
\boldsymbol{M}_{a(i)}^{(t)} = \pi \boldsymbol{M}_{r(i)}^{(t)} + (1 - \pi) \boldsymbol{\overline{M}}^{(t-1)},    
\end{equation*}
with the same weights $(\pi, 1 - \pi)$ as those used to compute the base population microbiota.\\ 

\paragraph{\textit{Environmental fixed effects}}

Taxa abundances can be modulated by environmental fixed effects, which may come from different sources and must be modeled accordingly. In particular, some covariates may not impact all taxa, individuals in a generation, or generations in the same way. To incorporate such effects, RITHMS incorporates the term $\boldsymbol{\theta} (\boldsymbol{E}_i^{(t)})^T$ to indicate the taxa, individuals, and generations for which environmental effects are applied from G1 onwards. In practice, we recommend that nonzero $\boldsymbol{\theta}$ values be drawn from a standard normal distribution to ensure they have the same scale as other parameters.\\

\paragraph{\textit{Genetic modulation}}

It has been shown that some taxa are correlated, regardless of whether or not a taxonomic link exists between them \citep{wuGuildbasedAnalysisUnderstanding2021}. It is thus reasonable to assume that the genetic modulation of taxa has a clustered structure so as to mimic existing correlations between taxa. In particular, simulations should yield strongly positive correlations between taxa in the same cluster, and weak or even negative correlations between taxa in different clusters. These clusters are identified from the rarefied taxa counts of the base population microbiota data $\boldsymbol{M}^{(0)}$, using hierarchical clustering based on Bray–Curtis dissimilarities, as previously applied for taxa clustering \citep{aguilarmarinLowerMethaneEmissions2020,pangClusteringSpatiallyAssociated2023}; note that other distance measures could be considered. By default, 100 clusters are generated. To choose the $\text{OTU}_\text{g}$ taxa (by default 5\% of all taxa) under genetic control, we randomly and iteratively select clusters of size 10 to 25 taxa until a threshold of $\text{OTU}_\text{g}$ taxa is reached.

The challenge then lies in constructing a sparse matrix of QTL effects on taxa CLR abundances, $\boldsymbol{\beta}$. The term $\boldsymbol{\beta} \boldsymbol{G}^{(t)}$ corresponds to the cumulative effect of the $\text{QTL}_\text{o}$ causative SNPs on the taxa. Here, by default we set $\text{QTL}_\text{o}$ to 20\% of the total number of SNPs ($n_\text{g}$) divided by the number of clusters under genetic control. To reach a given intra-cluster level of genetic correlation,  causative SNPs are sampled randomly for each cluster but common to all taxa within a cluster. The non-null coefficient $\beta_{sg}$ for genetically modulated taxon $s$ from cluster $c(s)$ with QTL $g$ is then set to $\beta_{sg} = \tilde{\beta}_{c(s)g} + \tilde{\beta}_{sg}$, where $\tilde{\beta}_{c(s)g}\sim \mathcal{N}(0, \sigma^2_{\beta})$ and $\tilde{\beta}_{sg} \sim \mathcal{N}(0, \sigma^2_{\beta})$. Note that $\tilde{\beta}_{c(s)g}$ depends only on the cluster, ensuring within-cluster correlation, whereas $\tilde{\beta}_{sg}$ is specific to each taxon under genetic control. In this way, as there may be some overlap in QTLs selected for different clusters, both intra- and inter-cluster genetic correlations are induced. The strength of this correlation is mainly limited by the number of clusters and the level of overlap of causative SNPs between clusters. The direction of correlation between taxa in two clusters is given by the sign of $\tilde{\beta}_{c(s)g} \times \tilde{\beta}_{c(s^\prime)g}$, summed over the common SNPs between the two clusters , as illustrated in the schematic diagram in Figure \ref{figure:graphical_wf}. Finally, for each taxon we center $\boldsymbol{\beta} \boldsymbol{G}^{(t)}$ to ensure both positive and negative genetic modulation, rather than systematic enrichment or depletion within the population.

We have provided a function within RITHMS, \texttt{calibrate\_gen\_effect()}, to help users evaluate the impact of $\sigma_{\beta}$ on the distribution of taxa heritabilities. In practice, it is often reasonable to expect the majority of taxa to have heritabilities on the order of 0.1, with maximum values of no more than 0.5 \citep{zangHeritableNonheritableRumen}.\\

\paragraph{\textit{Quantifying microbiota diversity}} 

A variety of metrics exist to quantify the $\alpha$-diversity (intra-sample diversity) from microbiota data. For a composition $\mathbf{p} = (p_1, \dots, p_{n_\text{b}})$, where $\sum_{1}^{n_\text{b}} p_j = 1$, we consider the Shannon index $H^1(\mathbf{p})$ defined as $H^1(\mathbf{p}) = - \sum_{j=1}^{n_\text{b}} p_j \log(p_j)$ with the convention that $0 \log(0) = 0$. When computed directly from sequencing data, this index is based on species counts transformed to relative abundances and thus suffers from potential undersampling of rare species. In order to obtain counts from the relative abundances and mimick this sampling step, $n_{\textrm{ind}}$ multinomial samplings $M(10000,(\mathbf{p_{i,1}},... \mathbf{p_{i,n_\text{b}}}))$ are performed, with $(\mathbf{p_{i,1}},...,\mathbf{p_{i,n_\text{b}})}$ the relative abundances of taxa for individual $i$, equivalent to the cutoffs on sequencing depth used in the dataset analyses described above. Note that 10 000 corresponds to the average sequecing depth in our motivating example but can be replaced with any value, \emph{e.g.} the mean depth in the base population, or vector of values, \emph{e.g.} samples from a sequencing depth distribution, whether synthetic (Gaussian, negative binomial, etc) or non-parametric (values in the base population). Numerous diversity indices have been proposed in the literature \citep{cazzollagattiEstimatingComparingBiodiversity2020,chaoUnifyingSpeciesDiversity2014a}, each capturing a distinct ecological specificity of the microbiota structure. While we focus here on the Shannon index as a popular index to account for differences in abundances, alternative metrics could be considered.

\subsubsection*{Transgenerational simulation of phenotypes} 

Phenotypes at generation $(t)$ are simulated as the result of the combined effects of the microbiota and direct genetic effects following the model developed by \citet{perez-encisoOpportunitiesLimitsCombining2021}:

\begin{equation}
\label{eq:phenotype}
\boldsymbol{y}^{(t)} = \boldsymbol{\alpha}^T \boldsymbol{G}^{(t)} + \boldsymbol{\omega}^T \boldsymbol{B}^{(t)} + \boldsymbol{\epsilon}_y^{(t)},
\end{equation}

\noindent with:
\begin{itemize}[itemsep=0pt,parsep=0pt]
\item $\boldsymbol{\alpha}$ the regression coefficients corresponding to the QTL effects on the phenotype ($1 \times n_\text{g}$),
\item $\boldsymbol{G}^{(t)}$ the genotype values of all individuals at generation $t$ ($n_\text{g} \times n_{\text{ind}}$),
\item $\boldsymbol{\omega}$ the regression coefficients corresponding to taxa effects on the phenotype ($1 \times n_\text{b})$,
\item $\boldsymbol{B}^{(t)}=\clr(\boldsymbol{M}^{(t)})$, the CLR-transformed relative abundance values for taxa of all individuals at generation $t$ $(n_\text{b} \times n_\text{ind})$,
\item $\boldsymbol{\epsilon}_y^{(t)} \sim \mathcal{N}(0, 1)$, univariate Gaussian noise.
\end{itemize}

\noindent Note that the variance of the Gaussian noise is set to $1$ to ensure that changes in  mean phenotypic values are expressed in units of standard deviations. 
In our simulation settings, we assume that all heritable taxa also have an effect on the phenotype; as such, the microbiota effect, corresponding to $\boldsymbol{\omega}^T \boldsymbol{B}^{(t)}$, also includes an indirect genetic effect.\\

\paragraph{\textit{Breeding values and heritability}} 

Under this formulation, we define the Direct Breeding Value as $\textbf{BV}_d^{(t)} = \boldsymbol{\alpha}^T \boldsymbol{G}^{(t)}$, the Microbiota-mediated Breeding Value as $\textbf{BV}_m^{(t)} = \boldsymbol{\omega}^T \boldsymbol{\beta G}^{(t)}$ and the Total Breeding Value 
$\textbf{BV}_t^{(t)} = \textbf{BV}_d^{(t)} + \textbf{BV}_m^{(t)}$.
This $\textbf{BV}_t^{(t)}$, containing all of the genetic effects on the phenotype, takes into account both the direct genetic effect ($\boldsymbol{\alpha}^T \boldsymbol{G}^{(t)}$) due to the transmission of the genotype and the indirect microbiota-mediated ones ($\boldsymbol{\omega}^T \boldsymbol{\beta G}^{(t)}$), due to the fraction of the microbiota that has an effect on the phenotype and is under genetic control.

From these quantities, it is possible to define a few quantities of interest: (1) the total heritability $h^2 = \left[ \text{var}\left(\boldsymbol{\alpha}^T \boldsymbol{G}^{(t)} + \boldsymbol{\omega}^T \boldsymbol{\beta G}^{(t)}\right) \right] \big/ \text{var}(\boldsymbol{y}^{(t)})$, (2) the direct heritability\\ $h^2_d = \text{var}(\boldsymbol{\alpha}^T \boldsymbol{G}^{(t)})/\text{var}(\boldsymbol{y}^{(t)})$, and (3) the microbiability $b^2 = \text{var}(\boldsymbol{\omega}^T \boldsymbol{B}^{(t)})/\text{var}(\boldsymbol{y}^{(t)})$.\\

\paragraph{\textit{Parameter calibration}}
\label{sec:calibration}The regression vectors $\boldsymbol{\alpha}$ and $\boldsymbol{\omega}$ are fixed across generations and calibrated on the base population. The calibration consists in rescaling $\boldsymbol{\omega}$ based on $\boldsymbol{\alpha}$ in order to reach user-specified values for the direct heritability $h^2_d$ and the microbiability $b^2$. If $h^2_d$ is set to 0, then all $\boldsymbol{\alpha}$ coefficients are set to zero and the calibration only affects $\boldsymbol{\omega}$. 
Initial values $\tilde{\boldsymbol{\alpha}}$ for the non-zero coefficients of $\boldsymbol{\alpha}$ are sampled from a $\Gamma(0.4, 5)$ distribution and $\tilde{\boldsymbol{\omega}}$ for the non-zero coefficients of $\boldsymbol{\omega}$ are sampled from a $\Gamma(1.4, 3.8)$, as done in the Simubiome method \citep{perez-encisoOpportunitiesLimitsCombining2021}, before rescaling takes place.\\

\paragraph{\textit{Selection}}

\label{sec:pheno} To select the individuals that will make up the breeding stock for the next generation, by default 30\% of the males and 30\% of the females are selected at each generation. This proportion reflects common practice in many breeding programs, as typical selection rates generally range between 20 \% and 40 \%, corresponding to a moderate selection intensity. These fractions can be modified via the parameters \texttt{size\_selection\_F} and \texttt{size\_selection\_M}. If no selection is specified, individuals are chosen at random. Otherwise, a user-specified selection criterion is used to rank individuals, and only a fraction (specified above) of the top performers are retained to reproduce and contribute to the next generation. The available selection criteria are:

\begin{itemize}[itemsep=0pt,parsep=0pt]
\item the microbiota effect, $\boldsymbol{\omega}^T \boldsymbol{B}^{(t)}$, measuring the contribution of the microbiota to the phenotype at generation $t$,
\item the Microbiota-mediated Breeding value, $\textbf{BV}_m^{(t)}$, capturing the portion of the host’s genetic effects that modulate the phenotype via their impact on microbiota composition,
\item the Direct Breeding Value, $\textbf{BV}_d^{(t)}$, quantifying the direct effect of the host genotype on the phenotype,
\item the Total Breeding Value, $\textbf{BV}t^{(t)}$, corresponding to the overall genetic expectation of the phenotype, combining both direct and microbiota-mediated genetic effects (in the absence of environmental or transmission effects), corresponding to classic genomic selection,
\item the microbiota diversity, $\boldsymbol{\delta}^{(t)}$, computed as the Shannon diversity index,
\item a weighted index combining microbiota diversity and total breeding value, $\textrm{w}{\textrm{div}}\boldsymbol{\delta}^{(t)} + (1-\textrm{w}{\textrm{div}})\textbf{BV}t^{(t)}$, with $\textrm{w}{\textrm{div}}$ set by the user.
\end{itemize}

These quantities represent different components of the potential of an individual for transmitting traits to the next generation. The genetic breeding value captures the contribution of host genetics, whereas the microbial diversity value reflects the influence of the microbiota on phenotype. The combined metric integrates both sources of variation, taking into account the idea that host performance may rely on both genetics and microbial composition.

\section*{Results}

In this section, we will explore a wide range of simulation scenarios to demonstrate the capabilities and features of RITHMS, from sanity checks to our initial use cases. These results were obtained on the dataset described in the \textit{Real hologenomic data used as a base population} section. Unless otherwise specified, all scenarios use the following simulation parameters: $h^2_d = 0.25$, $b^2 = 0.25$, $\sigma_{\beta}\times\sqrt{\text{QTL}_\text{o}}=0.3$, $\sigma_m =0.6$, $n_{\text{ind}} = 500$, $\lambda = 0.5$ and $n_{\text{gen}}=5$. Other parameters are set to default values as described in the package documentation.

\subsection*{Simulated microbiota reflect expected structure} 

\begin{figure*}[h]
\centerline{\includegraphics[width=\textwidth]{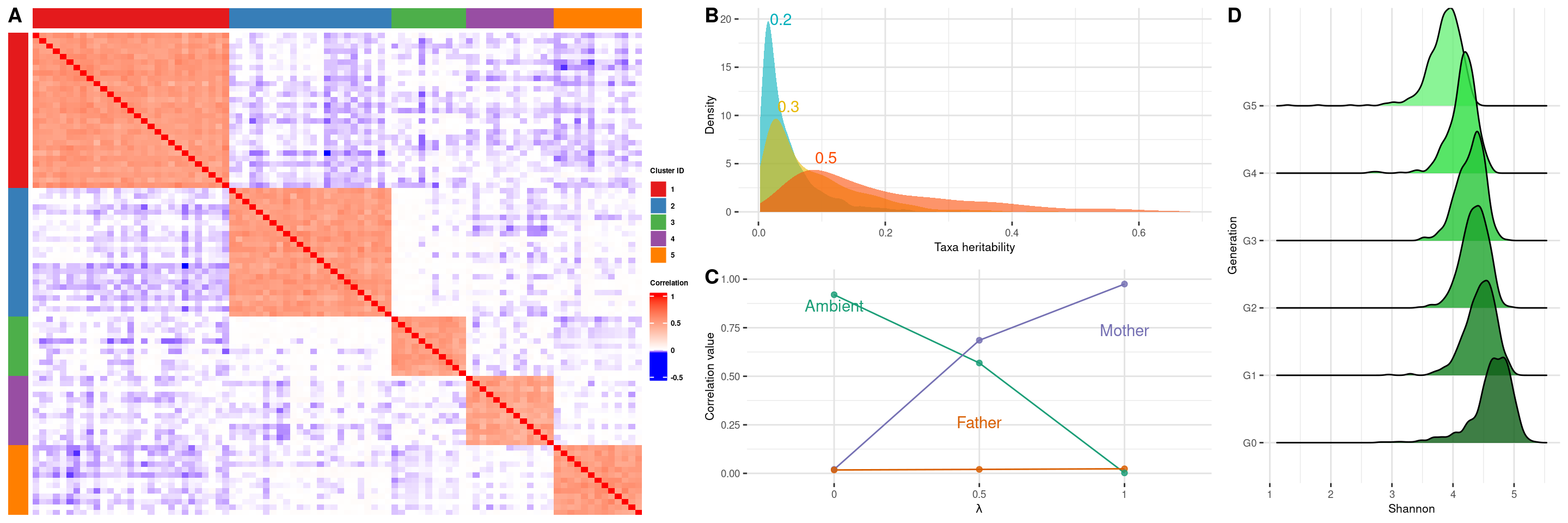}}
\caption{\label{figure:microbiome} Key characteristics of microbiota data simulated with RITHMS. (A) Pairwise correlation matrix of taxa abundances. Abundances were simulated assuming all taxa are under genetic control and distributed in five clusters (shown with color bars in the margins). Taxa are sorted based on the cluster they belong to. (B) Density plot of the distribution of taxa heritability for increasing genetic effect sizes ($\sigma_{\beta}\times\sqrt{\text{QTL}_\text{o}}$), shown above each curve.
(C) Correlation between offspring $\alpha$-diversity (from G2) and that of its mother (purple), father (orange) or ambient microbiota (green) for increasing values of $\lambda$. Correlations are computed from a population of 500 offsprings and averaged over 10 repetitions. (D) Density plots of the distribution of $\alpha$-diversity values in the base population (G0) and five consecutive generations (G1 to G5), in the absence of selection and environmental filters.}
\end{figure*}

We first evaluate whether the simulated microbiota exhibit expected characteristics. The pairwise correlation matrix of simulated abundances (Figure~\ref{figure:microbiome}A) shows that RITHMS successfully produces both strong intra-cluster genetic correlations as well as more modest inter-cluster anti-correlations, thanks to the set of partially overlapping QTLs between clusters. Likewise, increasing QTL effect sizes on taxa increases the heritability of taxa abundances, as expected (Figure~\ref{figure:microbiome}B). The density plots are produced by \texttt{calibrate\_gen\_effect()} and are intended to guide the user in choosing an appropriate effect size to achieve a target distribution of taxa heritabilities. In this setting, a reasonable distribution of taxa heritabilities appears to roughly correspond to a value of $\sigma_{\beta}\times\sqrt{\text{QTL}_\text{o}}=0.3$. We also confirm the impact of $\lambda$ in modulating the relative importance of vertical and horizontal transmission (Figure~\ref{figure:microbiome}C). When $\lambda = 0$, corresponding to no vertical transmission, offspring $\alpha$-diversity is strongly correlated with that of the ambient microbiota (values averaged over 10 simulated datasets). As $\lambda$ increases, so does the correlation between maternal and offspring microbiota $\alpha$-diversity. In constrast, the correlation between paternal and offspring microbiota diversity is low for all values of $\lambda$. This is expected, as the sire microbiota does not directly contribute to that of its offspring. Finally, in the absence of selection or environmental filters, the distribution of $\alpha$-diversity remains stable across generations (Figure~\ref{figure:microbiome}D), as expected for communities evolving in a neutral framework.

\subsection*{Introduction of sporadic or sustained environmental effects}

\begin{figure*}[h]
\centerline{\includegraphics[width=40pc]{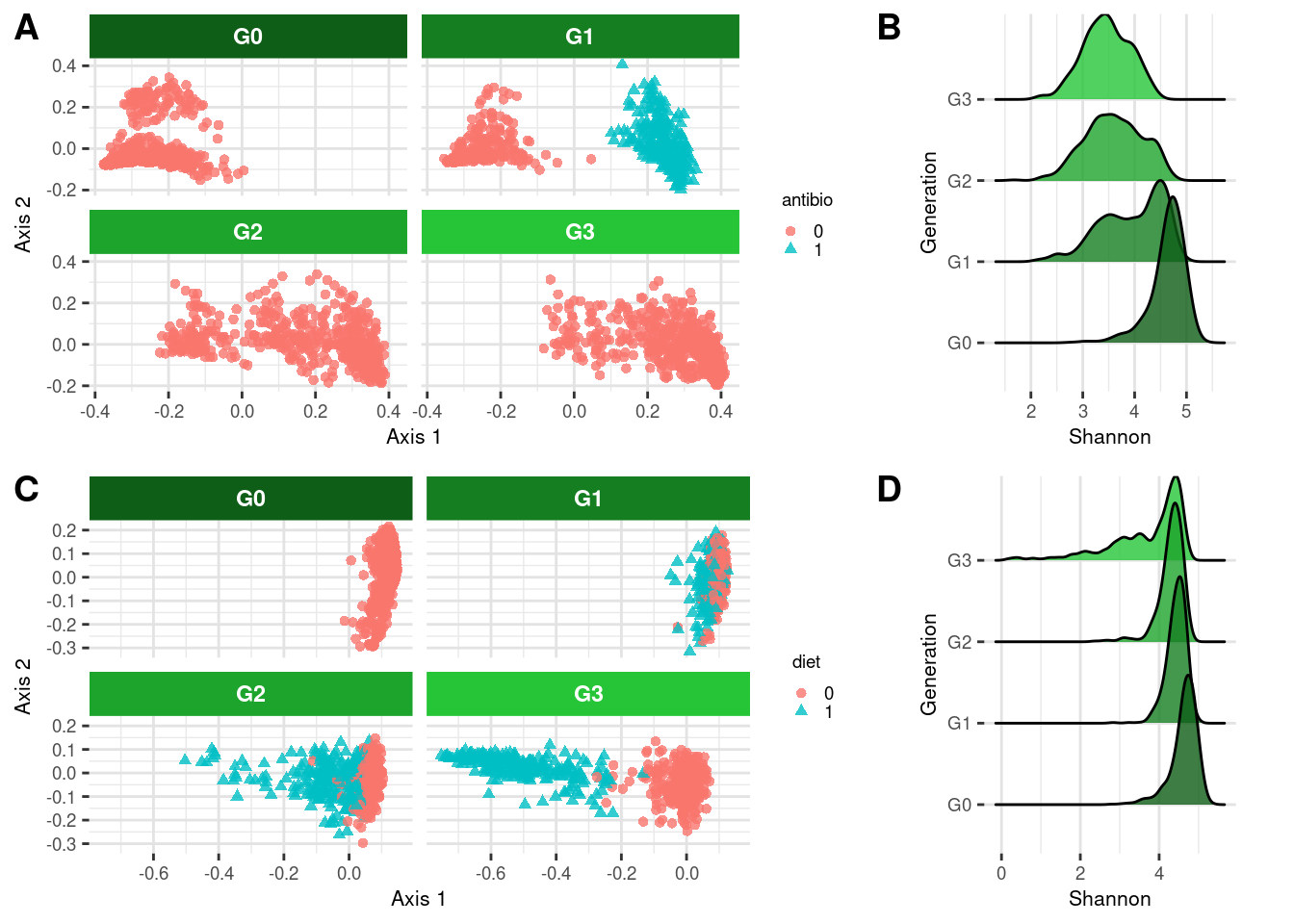}}
\caption{\label{figure:env}Simulation of sporadic (top) and sustained (bottom) environmental effects in RITHMS. (A) Multidimensional scaling (MDS) of microbial abundance data (Bray-Curtis distances). Half the individuals at G1 (blue triangles) are subject  to a sporadic antibiotic treatment. (B) Density plots of $\alpha$-diversity values before (G0), during (G1) and after (G2 to G3) sporadic antibiotic treatment. (C) Multidimensional scaling (MDS) of microbial abundance data (Bray-Curtis distances). Starting from G1, half the individuals at each generation (blue triangles) are subject to a diet favoring two clusters of taxa.(D) Density plots of $\alpha$-diversity values before (G0) and during (G1 to G3) sustained diet intervention.
}
\end{figure*}

In breeding and selection programs, it is essential to account for fixed environmental effects, given their strong role in modulating an individual's phenotype.
It is therefore important to verify that simulated transgenerational hologenomic data can correctly integrate such factors under a variety of plausible scenarios, such as short-term treatments or long-term diet effects. For the microbiota, as fixed environmental effects can be cumulated with varying effects on each taxa, RITHMS allows users to specify a (potentially sparse) $\boldsymbol{\theta}$ matrix, corresponding to the environmental effect sizes on CLR-transformed taxa abundances. To illustrate this, we consider two scenarios introducing either a sporadic (Figure~\ref{figure:env}A-B) or sustained (Figure~\ref{figure:env}C-D) environmental effect, as would respectively be the case if a subset of individuals in one generation were administered antibiotics or if individuals in each generation were randomly assigned to different diet groups. 

In the first case, we assume that half of the individuals in G1 are administered an antibiotic, provoking significant abundance changes across all taxa. The values for this effect were sampled from a normal distribution $\mathcal{N}(0, 5)$. This one-time environmental effect leads to a strong separation into two groups with very distinct microbiota compositions (Figure~\ref{figure:env}A) and constrasted $\alpha$-diversity, as evidenced by the bimodal distribution of $\alpha$-diversity values in generation G1 (Figure~\ref{figure:env}B). In the absence of continued antibiotic intake after G1, the lower diversity observed for the antibiotic group is progressively attenuated in the following generations due to random mating, and the bimodality disappears, although the $\alpha$-diversity is reduced on average compared to the base generation (\emph{e.g.}, when comparing G3 and G0 in Figure~\ref{figure:env}B). Likewise, the strong group structure in microbiota compositions induced by the treatment progressively disappears in following generations, but the diversity of the overall population shifts towards that of the antibiotic-treated microbiota, suggesting long-lasting changes of the treatment. Similar trends were observed when varying the $\lambda$ parameter, with stronger and longer-lasting antibiotic effects on diversity for higher values of $\lambda$ (results not shown).

In the second case, we assume that individuals from each generation following the base population are randomly assigned to one of two diets, one of which favors abundances in 2 randomly chosen taxa clusters. To simulate a relatively modest effect on the CLR-scale, non-zero values of $\theta$ were drawn from a normal distribution with smaller variance than that of the previous case, $\mathcal{N}(0, 2)$. This sustained environmmental effect induces a progressive separation of the diet groups that becomes particularly marked at G3 (Figure~\ref{figure:env}C). As two taxa clusters are preferentially favored in one of the diet groups, with the effect accumulating across generations, we remark the emergence of a group with an increasingly large drop in diversity (Figure~\ref{figure:env}D).

\subsection*{Impact of genomic, microbiota and hologenomic selection strategies}

\begin{figure*}[h]
\centerline{\includegraphics[width=40pc]{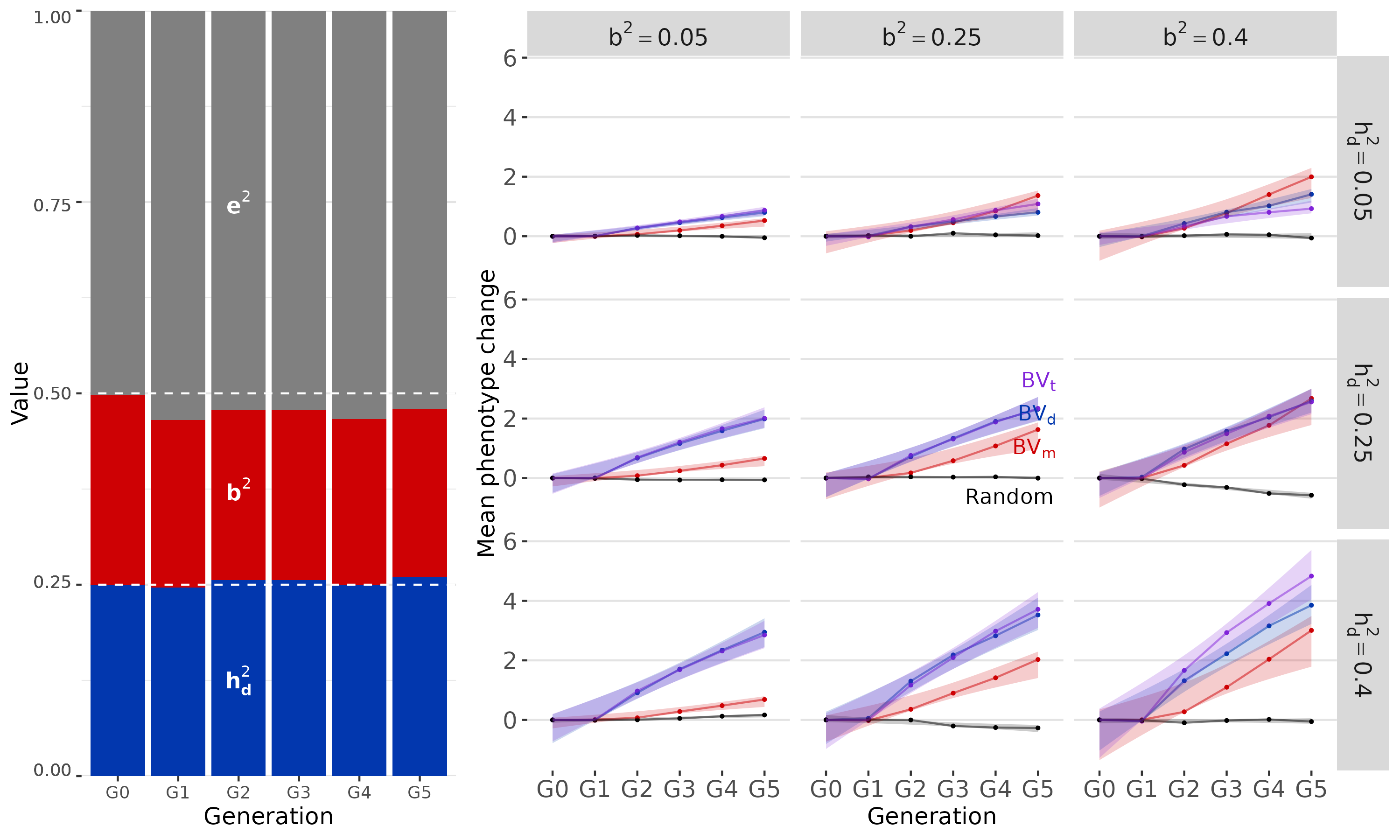}}
\caption{\label{figure:pheno}Direct heritability and microbiability of RITHMS simulations under various selection strategies. (A) Observed direct heritability $h_\text{d}^2$ and microbiability $b^2$ (averaged over 50 simulated datasets) in a scenario with random selection and target values $h_\text{d}^2 = b^2 = 0.25$.(B) Mean phenotypic change across five generations, (averaged over 50 simulated datasets, shaded regions correspond to 95\% confidence intervals) with $\lambda = 0.1$, according to various values of direct heritability (rows) and microbiability (columns) and different selection strategies: $\textbf{BV}_d^{(t)}$ (direct breeding values, blue line), $\textbf{BV}_m^{(t)}$ (microbiota breeding values, red line), $\textbf{BV}_t^{(t)}$ (total breeding values, purple line), random selection of parents for the next generation (black line).}
\end{figure*}

In the previous sections, we showed that the microbiota simulated by RITHMS reflect expected characteristics in terms of inter- and intra-cluster genetic correlations among taxa, taxa heritability, vertical or horizontal transmission, as well as microbiota diversity across generations, in the presence or absence of environmental effects. We now turn our attention to phenotypes simulated from the transgenerational hologenomic data under the model in Equation~\eqref{eq:phenotype}. Two critical user-provided parameters for RITHMS simulations are the direct heritability $h^2_d$ and microbiability $b^2$. In the absence of selection, we next sought to verify that the target values are reached and maintained across generations in the case of $h_\text{d}^2=b^2=0.25$ (Figure~ \ref{figure:pheno}A), corresponding to intermediate values and similar to those used in \citet{perez-encisoOpportunitiesLimitsCombining2021}. $h^2_d$ and $b^2$ were computed using the true values of $\boldsymbol{\alpha}$ and $\boldsymbol{\omega}$  and the simulated values of $\boldsymbol{G}^{(t)}$ and $\boldsymbol{M}^{(t)}$ at each generation. As $\boldsymbol{\alpha}$ and $\boldsymbol{\omega}$ are calibrated using the base population to achieve target heritability and microbiability (See the \textit{Parameter calibration} section), it is no surprise that $h^2_d$ and $b^2$ are exactly at $0.25$ for G0. In subsequent generations, the direct heritability varies only slightly around its target value, and we remark that the observed microbiability tends to be slightly lower than its target value. 

Given that the direct heritability and microbiability appear to be reasonably maintained near their target values in the absence of selection, we next evaluate trends in phenotypic improvement as a function of four different selection strategies for varying values of $h^2$ and $b^2$ (Figure~\ref{figure:pheno}B): selection of 30\% of males and 30\% of females based on (i) no criterion (random), (ii) the total breeding value ($\textbf{BV}_t^{(t)}$), (iii) the direct breeding value ($\textbf{BV}_d^{(t)}$), or (iv) the microbiota effect value. We observe that the phenotypic change is up to twice as large for higher values of direct heritability and microbiability ($h_\text{d}^2=b^2=0.4$) as compared to lower values ($h_\text{d}^2 =b^2=0.05$). Microbiota selection outperforms the other modes of selection only when microbiability is large compared to the direct heritability ($b^2=0.4$ and $h^2_d$=0.05). Generally speaking, given the modest contribution of vertical transmission used here ($\lambda = 0.1$) that affects only the female path of transmission, hologenomic selection appears to act as a form of mass selection and provides little selection gain compared to genomic selection alone. As an indication, these results were obtained based on a total of 1800 simulated datasets (4 selection modes $\times$ 9 pairs of $h^2_d$ and $b^2$ values $\times$ 50 repetitions for each), using the pig hologenomic data described above as a base population, corresponding to 770 minutes of computational time with a maximum memory usage of around 1GB RAM on a laptop with 16 GB RAM (Intel(R) Core(TM) i5-1135G7 CPU @ 2.40GHz x 8). An implementation for parallelizing repeated simulations is available and demonstrated in the package vignette.

\subsection*{
Constructing an index for mixed-objective selection}

To demonstrate the flexibility and usefulness of RITHMS, we consider a practical case study of a simplified
breeding program with a multi-trait objective: maximizing phenotypic change, based on a quantitative trait of interest $\mathbf{y}^{(t)}$, while preserving microbial $\alpha$-diversity. One way to achieve this is to use a selection score that combines both objectives into a single value. 
With access to hologenomic data at each generation, such a score can be constructed as a weighted combination of phenotypic change and diversity.  Formally, we define our selection index as $\textrm{w}_{\textrm{div}} \cdot\boldsymbol{\delta}^{(t)} + (1-\textrm{w}_{\textrm{div}})\cdot\textbf{BV}_t^{(t)}$ (see the \textit{Selection} section) as a linear combination of the microbial diversity $\boldsymbol{\delta}^{(t)}$ and the total breeding value $\textbf{BV}_t^{(t)}$, with weight $\textrm{w}_{\textrm{div}} \in [0, 1]$. Note that $\textrm{w}_{\textrm{div}} = 0$ corresponds to classic genomic selection. This index is used to identify the 30\% of males and 30\% of females constituting the breeding stock for the next generation.

\begin{figure}[h]
\centerline{\includegraphics[width=400pt]{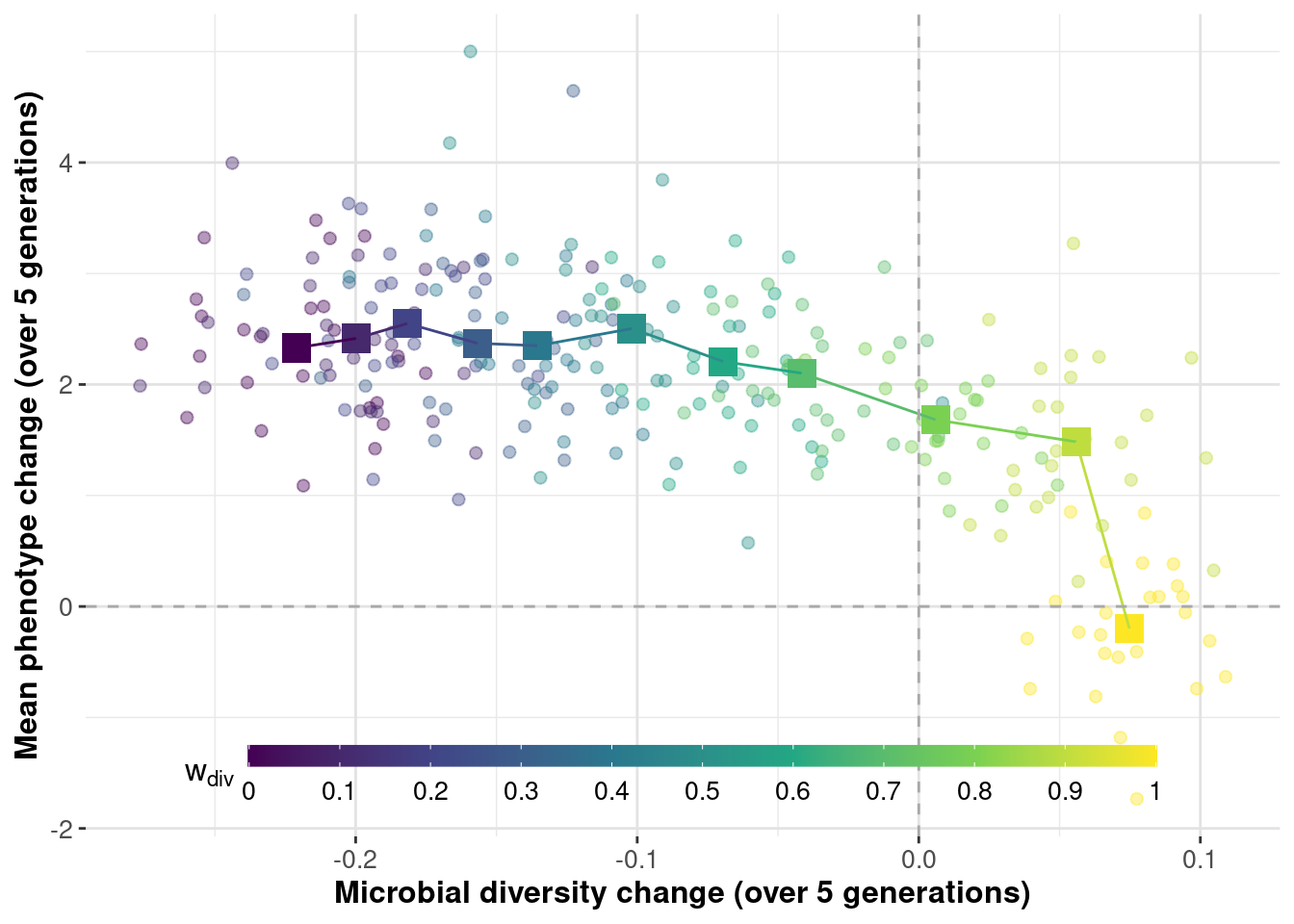}}
\caption{\label{figure:casestudy}Simulation-guided exploration of mixed selection index. Mean phenotype and microbial diversity changes from the base population (G0) to G5 as a function of $\textrm{w}_{\textrm{div}}$. The simulation is repeated 25 times for each value of $\textrm{w}_{\textrm{div}}$. Each simulation is shown as semi-transparent dots whereas square dots correspond to the mean computed over the 25 repetitions.}
\end{figure}

Here, we leverage RITHMS to construct a simulation study to identify an optimal weight to achieve gains on both components in a reasonable number of generations. In particular, we simulated data over five generations to evaluate the impact of $\textrm{w}_{\textrm{div}} \in \lbrace 0, 0.1, \ldots, 1\rbrace$ on changes in microbial diversity and phenotypic change, with direct heritability $h^2_d$ = 0.25, microbiability $b^2$ = 0.25, vertical transmission $\lambda$ = 0.5 and $n_{\text{ind}} = 500$ individuals per generation (Figure~\ref{figure:casestudy}). Although there is considerable variability among simulated datasets, we remark that there is a tradeoff between mean phenotypic change and microbial diversity (\emph{i.e.}, one comes at the expense of the other), which varies with $\textrm{w}_{\textrm{div}}$. Larger weights ($\textrm{w}_{\textrm{div}}$ = 0.8 or 0.9) simultaneously achieve phenotypic improvement and increased microbial diversity after five generations. However, for these scenarios, phenotypic change is more modest than for scenarios that increasingly mimic classic genomic selection ($\textrm{w}_{\textrm{div}} < 0.8$). These results suggest that a value of $\textrm{w}_{\textrm{div}} = 0.6$ achieves phenotypic change comparable to classic genomic selection in this case study, while drastically limiting the loss of microbial diversity.

\subsection*{Impact of data subsampling on variance component estimation}

\begin{figure}[h]
\centerline{\includegraphics[width=\textwidth]{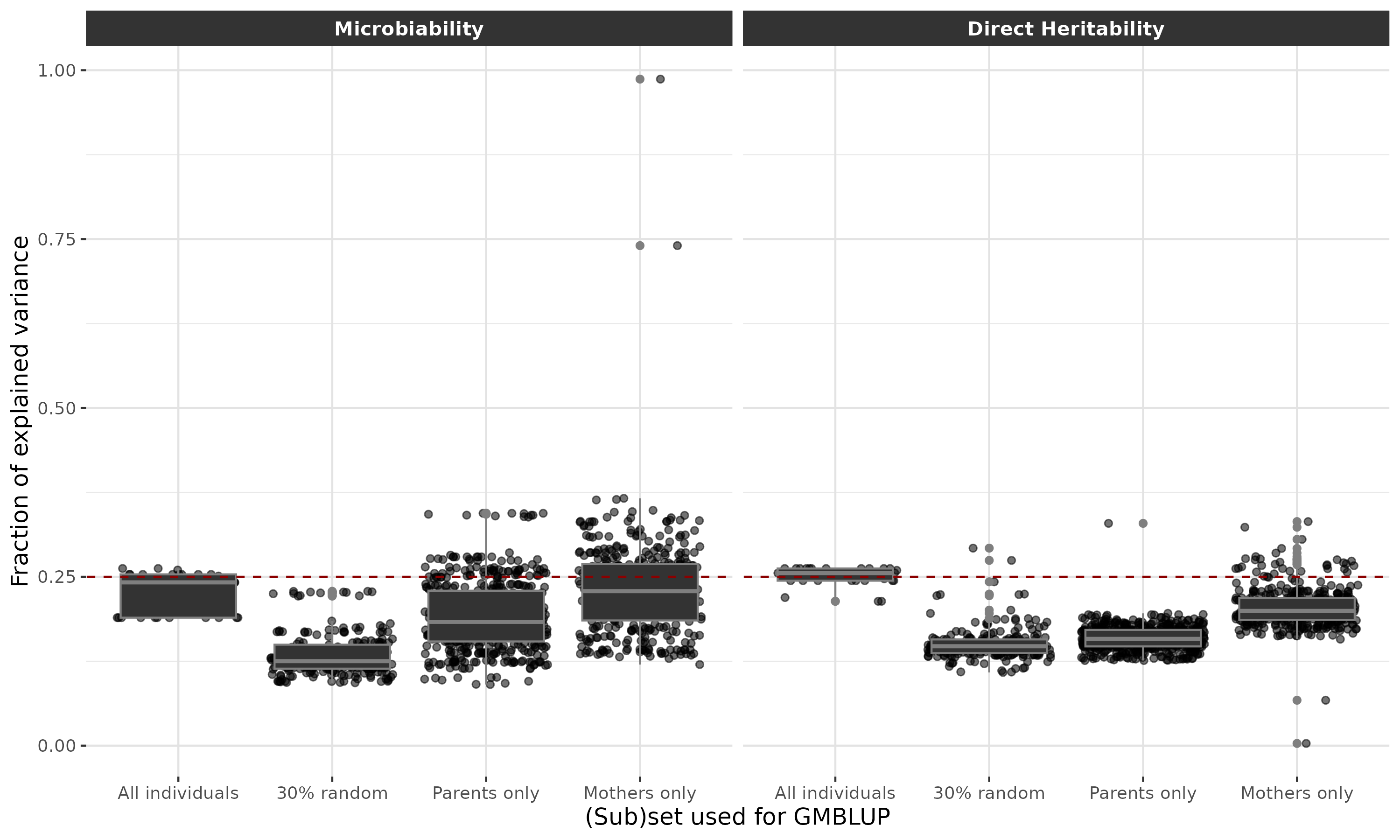}}
\caption{\label{figure:GMBLUP}Estimated variance components for microbiability ($b^2$) and direct heritability ($h^2_d$) obtained using GMBLUP over 500 simulation replicates under different subsampling scenarios: all individuals (2500 individuals), 30\% random (750 individuals), parents only (750 individuals), and mothers only (375 individuals). \review{The red dotted line corresponds to real simulated values ($b^2 = h^2_d = 0.25$).}}
\end{figure}

To illustrate another possible use of RITHMS, we conducted a study of the estimation of variance components under different data availability scenarios. Such a study is of particular interest to evaluate the impact of different sampling strategies, for example to account for financial or experimental constraints. First, we simulated a population of 2500 hologenotyped and phenotyped individuals over 5 
generations (500 per generation), with
$\lambda = 0.5$. 
Then we estimated the direct heritability ($h^2_d$) and microbiability ($b^2$) with a GMBLUP model using the \textit{BGLR} R package \citep{perezGenomewideRegressionPrediction2014} and linear scaled kernels of genotypes and microbiota, based on various subsets of individuals: (a) the full dataset (2500 individuals), (b) a random subset of 30\% of individuals (750 individuals), (c) only breeding parents (750 individuals), and (d) only mothers (375 individuals). Figure~\ref{figure:GMBLUP} shows the distribution of the estimated variance components across 500 independent iterations for each subset.

As expected, the most accurate estimates were obtained when all individuals were hologenotyped. Random subsampling led to a systematic underestimation of both $h^2_d$ and $b^2$, reflecting the loss of information about the underlying genetic and microbial covariance structures. When restricting the analysis to mothers only, estimates remained close to those obtained with the full dataset, even though a substantial part (85\%) of the information was omitted. This reflects the key role played by maternal microbiomes for phenotypic variance decomposition, notably due to the vertical transmission of the microbiota through mothers only. These results highlight the importance of a  meaningful subsampling strategy for hologenomic selection, e.g. focusing on reproducing individuals, to mitigate the loss of accuracy associated with incomplete data collection. An interesting follow-up study could investigate cases where genotypes are collected on a large number of individuals and microbiota data on only a targeted subset;
using all available genotypes, even when microbiome data are missing, would represent a more realistic setting and powerful setup for future applications.

\section*{Discussion}

In this work, we introduced a novel algorithm for simulating transgenerational hologenomic data, implemented in the R package RITHMS. Our tool expands the scope of existing genomic simulation methods \citep{pookMoBPSModularBreeding2020, gaynorAlphaSimRPackageBreeding2021} by adding a microbiota compartment and of existing hologenomic simulation methods \citep{perez-encisoOpportunitiesLimitsCombining2021} by enabling the simulation of multiple generations. In contrast to the only other transgenerational hologenomic simulator currently available, HoloSimR \citep{casto-rebolloHoloSimRComprehensiveFramework2025}, RITHMS uses real data as input, structures the microbiota into taxa clusters and incorporates potential environmental covariates. RITHMS directly accounts for the structure and characteristics of the microbiota as well as its complex transmission mechanisms (from both the mother and the ambient environment, with filters linked to host genetics) and the impact of sporadic or sustained environmental covariates. It is possible to calibrate both (i) the size of genetic effects on the microbiota to obtain a realistic distribution of taxa heritability and (ii) the direct genetic and microbial effects to achieve target values of direct heritability and microbiability. Complex breeding schemes using the genome, the microbiota or the hologenome combined with different selection scores or sampling strategies were used to showcase the flexibility and usefulness of RITHMS. RITHMS is available as an R package, runs on a commercial laptop and is able to generate transgenerational hologenomic data ($n_\text{g}=5000$, $n_\text{b}=2000$ taxa, $n_{\text{ind}}=500$ individuals) for five generations  in a few seconds. 

Our approach presents several limits and opportunities for future improvements. First, we remark on the slight negative bias we observed between the simulated and target values for microbiability $b^2$ (Figure~\ref{figure:pheno}A, from generation G1 onwards). Since taxa effects on the phenotype $\boldsymbol{\omega}$ are calibrated on G0, we hypothesize that this bias originates from a small loss of $\alpha$-diversity between G0 and G1, as the model cannot reproduce fully the complexity of the base population. Second, our simulation framework is based on a linear model, which has the advantage of being both interpretable and computationally tractable; however, it would be of interest to explore alternatives such as neural networks to introduce non-linearity into RITHMS. Third, our simulated microbiota correspond to snapshots in the lifetime of an animal that are intended for use in predictive models of hologenomic breeding values.
However, the microbiota corresponds to a highly dynamic measure that evolves throughout an animal's life, and future work could consider a dynamic model to simulate the microbiota at different time points. 

Likewise, it would be interesting to extend the RITHMS model to (i) account for microbial interactions with a non-diagonal covariance matrix for the noise component $\sigma_{m}$ of the taxa abundances, (ii) allow for the inclusion of more complex environmental effects, and (iii) allow for the use of semi-complete, rather than fully paired, genomic and microbiota data to create the base population, which would enable RITHMS simulations to be calibrated on a datasets for which some samples lack genomic or microbiota data.

Finally, in future work we plan to extend the use of RITHMS to alternative hologenomic datasets, notably for a variety of species and experimental designs, and additional use cases for the evaluation of complex breeding schemes.

\section*{Acknowledgements}

The authors gratefully acknowledge helpful discussions with Thierry Tribout and Vanille Déru and the breeding companies Axiom and Nucleus for providing animals through France Génétique Porc. Finally, the authors thank the UE3P phenotyping station staff in Le Rheu for animal raising and data recording (\url{https://doi.org/10.15454/1.5573932732039927E12}). Preprint version 4 of this article has been peer-reviewed and recommended by Peer Community In PCI Math Comp Biol (\url{https://doi.org/10.24072/pci.mcb.100417}; \cite{galiezToolHelpDesign2025}).

\section*{Fundings}

This research is supported by the French National Research Agency under the France 2030 program ("ANR-22-PEAE-0006"), which supports future innovation in agricultural science. 

\section*{Conflict of interest disclosure}

The authors declare that they comply with the PCI rule of having no financial conflicts of interest in relation to the content of the article.

\section*{Data, script and code availability}

RITHMS is an open-source package available on GitHub (\url{https://github.com/SolenePety/RITHMS}) and all data and code used for this study are available in Zenodo (\url{https://doi.org/10.5281/zenodo.17629188}; \cite{solenepety_2025_17629188}). A subset of the original dataset is available and can be used to reproduce the figures of the paper in a vignette.

\printbibliography
\end{document}